
%
%
%
\input mtexsis
\def\lmargin#1{\advance\hoffset by #1
                        \advance\hoffset by -1in
                        \advance\hsize by -\hoffset}
\def\rmargin#1{\advance\hsize by 1in
                        \advance\hsize by -#1}
\def\tmargin#1{\advance\voffset by #1
                        \advance\voffset by -1in
                        \advance\vsize by -\voffset}
\def\bmargin#1{\advance\vsize by 1in
                        \advance\vsize by -#1}
\def\icfp#1{\bigskip\centerline{Int.~Class.~for~ Physics: #1}\smallskip}
\input epsf
\def\readfigfile#1#2{\centerline{\epsfysize=#2 truein\epsffile {#1} }}
\def\d{{\rm d}}
\def\e{{\rm e}}
\def\braket#1#2{\langle #1 \big\vert #2 \rangle}
%
\lmargin{40mm}
\rmargin{20mm}
\Eurostyletrue
\superrefsfalse
\preprint
\pubcode{SAL-TH-94-02, {hep-th/9501108}}
\titlepage
\title
Massless Scalar Field Theory in a Quantised Space-Time
\endtitle
\authors
J.~C.~Breckenridge,$^a$ V. Elias,$^b$ and T.~G.~Steele $^a$
\institution{a}{Department of Physics and Engineering Physics and
Saskatchewan Accelerator Laboratory,
University of Saskatchewan, Saskatoon, Saskatchewan, Canada S7N 0W0}
\institution{b}{Department of Applied Mathematics, University of Western
Ontario, London, Ontario, Canada N6A 5B9}
\endauthors
\abstract
A  method is developed to construct a non-local massless scalar
field theory in a flat quantised space-time generated by an operator
algebra.
Implicit in the operator algebra is a fundamental length scale of the
space-time.  The fundamental two-point function of free fields is
constructed
by assuming that the causal Green functions still have support on the
light cone in the operator algebra quantised space-time.
In contrast to previous stochastic approaches, the method introduced here
requires
no explicit averaging over spacetime coordinates.
The two- and four-point functions of~$g \varphi^4$
theory are  calculated to the one-loop level, and no ultraviolet
divergences are encountered.   It is also demonstrated that there are no
IR divergences in the processes considered.
\endabstract
\icfp{0370, 1110, 04, 0460}
\endtitlepage
\referencelist
\reference{kato} M.~Kato,
\journal Phys. Lett.;245,43(1990)
\endreference
\reference{net} L.~J. Garay,
gr-qc/9403008,~Imperial/TP/93-94/20~(1994)
\endreference
\reference{pais} A.~Pais and G.~E.~Uhlenbeck,
\journal Phys. Rev.;79,145(1950)
\endreference
\reference{snyder1} H.~S.~Snyder,
\journal Phys. Rev.;71,38(1947)
\endreference
\reference{Lee} T.~D.~Lee,  \booktitle{Discrete Mechanics.} Erice:
Lectures at the International School of Subnuclear Physics, 1983
\endreference
\reference{snyder2} H.~S.~Snyder,
\journal  Phys. Rev.;72,68(1947)
\endreference
\reference{golfand1} Yu.~A.~Gol'fand,
\journal Soviet Journal JETP;37,504(1959)
\endreference
\reference{golfand2} Yu.~A.~Gol'fand,
\journal Soviet Journal JETP;43,256(1962)
\endreference
\reference{nams1} Kh.~Namsrai,
\journal Intern. J. Theor. Phys.;20,365(1981)
\endreference
\reference{nams2} Kh.~Namsrai,
\journal Intern. J. Theor. Phys.;24,741(1985)
\endreference
\reference{nams3} Kh.~Namsrai,\booktitle{ Nonlocal Quantum Field Theory
and Stochastic Quantum Mechanics.}  Dordrecht: Reidel, 1986
\endreference
\reference{nams4} Kh.~Namsrai,
\journal Fortschr. Phys.;36,479(1988)
\endreference
\reference{nams5} Kh.~Namsrai,
\journal Intern. J. Theor. Phys.;30,587(1991)
\endreference
\reference{namsdin} Kh.~Namsrai and M. Dineykhan,
\journal Intern. J. Theor. Phys.;22,131(1983)
\endreference
\reference{dinnams1} M.~Dineykhan and Kh.~Namsrai,
\journal Intern. J. Theor. Phys.;24,1197(1985)
\endreference
\reference{dinnams2} M. Dineykhan and Kh. Namsrai,
\journal Intern. J. Theor. Phys.;25,685(1986)
\endreference
\reference{dinnams3} M. Dineykhan and Kh. Namsrai,
\journal Intern. J. Theor. Phys.;28,719(1989)
\endreference
\reference{dinnams4} M. Dineykhan and Kh. Namsrai,
\journal Intern. J. Theor. Phys.;29,311(1990)
\endreference
\reference{dinefnams} M. Dineykhan, G.~V. Efimov, and Kh. Namsrai,
\journal Intern. J. Theor. Phys.;28,1463(1989)
\endreference
\reference{blok} D.~I. Blokhintsev, \booktitle{Space and Time in the
Microworld.} Dordrecht: Reidel, 1973
\endreference
\reference{grensing} G. Grensing,
\journal J. Phys. A;10,1687(1977)
\endreference
\reference{guersey} F.~Guersey, Introduction to Group Theory in
\booktitle{Relativity, Groups and Topology} Eds. C.~ DeWitt and B. DeWitt
\endreference
\reference{birdav} N.~D. Birrell and P.~C.~W. Davies, \booktitle{ Quantum
Fields in Curved Space.}  Cambridge:  Cambridge University Press, 1982
\endreference
\reference{PandT} P.~Pascual, R.~Tarrach, \booktitle{ QCD:
Renormalization for the Practitioner.},
New York: Springer-Verlag, 1989
\endreference
\reference{kreimer} D.~Kreimer,
\journal Z. Phys. C;54,667(1992)
\endreference
\reference{genun1} M.~Maggiore,
\journal Phys. Lett. B;319,83(1993)
\endreference
\endreferencelist

\offparens
\section{Introduction}

The existence of a minimum observable length in string theories \cite{kato}
and the evidence suggesting the existence of fundamental length scales in
quantum gravity
\cite{net}
suggests that the construction of a quantum field theory~(QFT)
in space-times which possess a fundamental scale becomes an important
question.
Manifestly non-local actions for QFTs contain derivatives of infinite
order and
necessarily contain fundamental length scales while retaining the
continuous nature of
spacetime\cite{pais}.

An alternative appproach
of the
formulation of  QFT in a (flat) quantised space-time was first attempted by
Snyder\cite{snyder1}. This quantised space-time naturally introduces a
fundamental length scale,
which as well as describing the space-time itself, can also regulate UV
divergences of the QFT.  This latter aspect was the motivation for
Snyder's work.

The Snyder space-time is constructed essentially by adding non-commuting
operators for the space-time coordinates to the usual Poincar\'e group.
The spatial coordinate operators have spectra that are integer multiples
of a fundamental
length $a$, while the time coordinate operator admits a continuous
spectrum. Momentum space
remains continuous, with commuting momentum operators.  These features
result in a space-time structure
violating the Born reciprocity principle, or the symmetry between
configuration
and momentum space representations of field theory.

 This lack of  reciprocity  makes the construction of field theories
difficult,
 due to   reliance on this feature of most of the presently established
formalisms.
 The common operations of analysis are either extremely cumbersome,
having to be carried out in
 terms of summations (see Lee\cite{Lee}), or cannot be carried out at
all.   Snyder\cite{snyder2}
 was able to formulate a version of the classical electromagnetic field
in his quantised
 space-time but  there is no evidence that this approach was effective
for the case of a quantum field theory.

 Gol'fand\cite{golfand1}\cite{golfand2}, advocated  the construction of a
quantum
 field theory in a momentum space of constant curvature.   This is in
some respects similar to
 the Snyder construction since the Snyder spacetime is generated by an
operator algebra
 formulated in an abstract space of constant curvature.  However, in
Gol'fand's theory,
 the addition of momenta are non-commutative, resulting in changes to the
laws of conservation of momentum and energy.

Attempts to restore symmetry between coordinate and momentum space, thus
maintaining
access to established methods of QFT are the basis of the work of
Namsrai\refrange{nams1}{nams5},Namsrai and Dineykhan\cite{namsdin},
Dineykhan and Namsrai\refrange{dinnams1}{dinnams4},
and  Dineykhan, Efimov and Namsrai\cite{dinefnams}.  These researchers
use the
concept of a stochastic space-time, in which coordinates are assumed to
undergo quantum fluctuations.
Explicit averaging over these stochastic coordinates removes the
asymmetry between configuration
and momentum space.  This allows Namsrai\cite{nams2} to make use of the
existing formalism of field
theory, with certain changes introduced by the averaging.

 The combination of the space-time stochasticity and the averaging
results in the theory
 having a manifestly  non-local nature, as a result of the explicit
dependence
 of the Green's functions, or propagators, on derivatives of infinite
order
 \cite{pais}\cite{blok}.  This approach  allows the familiar construction
of field
 theories in terms of an action principle with appropriately modified
propagators,
 which have the important property of being sufficiently convergent that
the resulting
 theory is ultraviolet finite.  The averaging over the space-time,
however, has the
 disadvantage of being rather ad hoc, and not contained within the
dynamics of the theory.

In this paper a new method will be used to construct a massless scalar
field
in a quantised space-time generated by an operator algebra. Section~2
develops the algebra of the space-time, and it is shown that the
Poincar\'e algebra remains an invariant subgroup.

The massless free scalar fields are considered in Section~3 leading to a
modified
free propagator, which depends on the fundamental scale of the
space-time.  The one-loop two-,
and four-point functions are calculated in Section~4 for massless
scalar~$g \varphi^4$
in the quantised space-time.
The free propagator is  compared to the propagator of the Namsrai theory
and
found to have  remarkably similar behaviour in the energy regime~$p^2 <
1/a^2$.
The large $p^2$ asymptotic behaviour, however, is quite different.  This
difference is perhaps not
surprising since failures of the stochastic averaging at small scales
where the algebra of the
quantized spacetime becomes crucial will lead to distinct large $p^2$
behaviour.


\section{Operator Algebra for a Quantised Space-Time}

\subsection{ The Snyder Algebra}

Here we present the  algebra for a quantised space-time as it was
developed by
Snyder\cite{snyder1}.  The algebra is formulated in an abstract
five-dimensional
space, which allows the construction of  the  Poincar\'e group, and
position operators~$x_ \mu$.

The Snyder algebra is based on the invariance of the homogeneous
quadratic form
$$
- \eta^2 = \eta_0^2 - \eta_1^2 - \eta_2^2 - \eta_3^2 - \eta_4^2 \EQN quadform
$$
in an abstract  $\eta$--space which is known as a de~Sitter space.  It is
known
that the symmetry group of a  de~Sitter space is  isomorphic to the
Poincar\'e
group in the limit of vanishing curvature~\cite{grensing}.  The
coordinate operators
are defined over this space as
$$
\eqalign{
x &= i a \left(\eta_4\, {\partial \over{\partial \, \eta_1}}- \eta_1 \,
{\partial \over{\partial \, \eta_4}}\right) \cr
y &= i a \left(\eta_4\, {\partial \over{\partial \, \eta_2}}- \eta_2 \,
{\partial \over{\partial \, \eta_4}}\right) \cr
z &= i a \left(\eta_4\, {\partial \over{\partial \, \eta_3}}- \eta_3 \,
{\partial \over{\partial \, \eta_4}}\right) \cr
t &= i {a \over{c}}\! \left(\eta_4\, {\partial \over{\partial \,
\eta_0}}+\> \eta_0 \,
{\partial \over{\partial \, \eta_4}}\right) \cr}\EQN coordops
$$
where~$a$ is the fundamental length introduced by this theory.  Rotation
operators, ie., angular momentum ($L$)
and boosts ($M$), are defined analogously to the conventional ones:
$$
\eqalign{
L_x &= i \hbar \left(\eta_3\, {\partial \over{\partial \, \eta_2}}-
\eta_2 \,
{\partial \over{\partial \, \eta_3}}\right) \cr
L_y &= i \hbar \left(\eta_1\, {\partial \over{\partial \, \eta_3}}-
\eta_3 \,
{\partial \over{\partial \, \eta_1}}\right) \cr
L_z &= i \hbar \left(\eta_2\, {\partial \over{\partial \, \eta_1}}-
\eta_1 \,
{\partial \over{\partial \, \eta_2}}\right) \cr
\noalign{\vskip5pt}
M_x &= i \hbar \left(\eta_0\, {\partial \over{\partial \, \eta_1}}+
\>\eta_1 \,
{\partial \over{\partial \, \eta_0}}\right) \cr
M_y &= i \hbar \left(\eta_0\, {\partial \over{\partial \, \eta_2}}+\>
\eta_2 \,
{\partial \over{\partial \, \eta_0}}\right) \cr
M_z &= i \hbar \left(\eta_0\, {\partial \over{\partial \, \eta_3}}+\>
\eta_3 \,
{\partial \over{\partial \, \eta_0}}\right) \cr}\EQN rotateops
$$

Each of these operators is a symmetry operation on the de~Sitter space,
which
leaves the quadratic form \Ep{quadform} invariant.  It can easily be seen
that
the coordinate operators do not commute.  Their commutators are
$[x,y] =(i\,a^2/\hbar)\,L_z$ and cyclic permutations, and those involving
time obey
$[t,x] = (i\,a^2/\hbar c)\, M_x$, etc.
(Hereafter we will put~$c = \hbar = 1$.)
It is also seen that in the limit $a\to 0$
all commutators return to those associated with conventional,
non-quantised spacetime.
This correspondence is regarded as essential in all following work.

In a covariant form, the algebra becomes
$$
\EQNalign{
\left[ x^\mu, x^\nu\right] &=i a^2J^{\mu\nu} \EQN cov; a\cr
\left[x^\mu,
J^{\rho\sigma}\right]&=i\left(g^{\mu\rho}x^\sigma-g^{\mu\sigma}x^\rho\right)
\EQN cov;b\cr
[J_{\mu \nu}, J_{\rho \sigma}] & = i \left(g_{\nu \rho} J_{\mu \sigma}
- g_{\mu \rho} J_{\nu \sigma} + g_{ \mu \sigma} J_{\nu \rho} - g_{\nu
\sigma} J_{\mu \rho} \right) \EQN cov;c
}
$$
where $J_{\mu\nu}=-J_{\nu\mu}$
($L_x=J_{23},~L_y=J_{31}~, M_x=J_{10}~ etc$)
satisfy the usual algebra of Lorentz transformations.
The algebra defined by  \Ep{cov} is actually the de Sitter group
which has the following two casimir invariant operators~\cite{guersey}
$$
\EQNalign{
C_1&=x^2-\frac{1}{2}a^2J_{\mu\nu}J^{\mu\nu} \EQN cas;a\cr
C_2=\sum_{i=0}^4 W_iW_i\quad &;\quad W_i=\epsilon_{ijklm}J^{jk}J^{lm}\EQN
cas;b
}
$$
where $J_{kl}=-J_{lk}$ is defined by
$$
\EQNalign{
J_{kl}&\equiv aJ_{\mu\nu}\quad ;\quad \mu=k,~\nu=l,~~k,l=0,1,2,3 \EQN
cas2;a\cr
J_{4i}&\equiv x_\mu\quad ;\quad \mu=i=0,1,2,3 \EQN cas2;b
}
$$

In addition to the ten operators defined above, to take the place of
momentum operators we define
$$
 p_\mu = {1 \over{a}}\, \left(\eta_\mu \over{\eta_4}\right)
 \EQN momentops
$$
which have the required property of being translation operators.   It is
worth
noting that with these definitions we will have~$[p_ \mu, p_ \nu] = 0$.
Also,
it can easily be verified that we can write~$L_x = yp_z - zp_y$, etc., in
the standard way.
This constitutes the complete set of objects in the Snyder algebra.

We note from equation \Ep{coordops} that the Snyder coordinate operators
are very
similar in structure to the usual angular momentum operators of quantum
mechanics.
This demonstrates the quantised nature of the spectra of these
operators. The time operator, however, can be shown to possess a
continuous spectrum
although since the time and space operators do not commute
simultaneous space and time eigenstates cannot exist.  This latter point
will be discussed
further in the following sections.

\subsection{ Covariant Form of  Snyder Algebra}

With the inclusion of momentum operators, the Snyder algebra can be
easily written in covariant form by  working out the
commutator~$[x, p_x]$, where~$x$ and~$p_x$ are as given in section one,
along with its companions:
$$
\eqalign{
[x,p_x] & =  i \left( 1 + a^2 p_x^2 \right)\cr
[t,p_t] & = i \left(1 - a^2p_t^2 \right),\cr
[x,p_y]& = [y,p_x] = i a^2 p_xp_y,\cr
[x,p_t]& = [p_x, t] = i a^2 p_x p_t,\cr
{\rm etc.}\cr}\EQN commute-2
$$
It is easy to see that we can write the entire set of commutators
\Ep{commute-2} in the compact form
$$
[x_ \mu,p^\nu] = i \left(\delta_ \mu{}^\nu - a^2 p_ \mu p^\nu
\right).\EQN commute-set
$$
With the set of commutators written in covariant form we are now in a
position
to write down by inspection a form for $x_ \mu$ which leads to
\Ep{commute-set}.  This is
$$
\tilde x_ \mu = i \left(\partial_ \mu - a^2 p_ \mu \left(p^\rho \partial_
\rho \right) \right) \EQN x-sol
$$
where
$$
\partial_ \mu \equiv { \partial \over{\partial p^\mu}}
$$
and $\tilde x_\mu$ is used to denote the momentum space representation of
the coordinate operator.
Unless otherwise stated, all differentiations will be with respect to
momentum variables.

To determine the generator $J_{\mu\nu}$ consider
$$
[x_ \mu, x^\nu] = i a^2 J_ \mu{}^\nu . \EQN j-commute
$$
A simple computation using \Ep{x-sol} gives
$$
[x_ \mu , x^\nu]  = i a^2 \left( p_ \mu x^\nu - p^\nu x_ \mu \right)
\EQN commute-verify
$$
which is exactly the desired form provided that $p_ \mu x^\nu - p^\nu x_
\mu$,
which has the operators in reverse order from what is usually written
down,
is the correct representation for $J_ \mu{}^\nu$ in momentum space.  That
this
form  is equivalent to the more usual  representation can be shown by
using equation~\Ep{commute-set}.
We prefer the form given above since it places the derivative operators,
contained
in $ x_ \mu$, to the right of the momentum variables.

\subsection{  Correspondence of the Snyder Algebra and Poincar\'e Group}

Straightforward manipulation will serve to verify the following:
$$
\EQNalign{
[p_ \mu, J_ {\rho \sigma}]  & = i( g_{\mu \rho} p_ \sigma - g_{\mu
\sigma} p_ \rho )\EQN corres-1;a \cr
[p^2, J_{\rho \sigma}]  &= 0 \EQN corres-1;b \cr
[J_{\mu \nu}, J_{\rho \sigma}] & = i \left(g_{\nu \rho} J_{\mu \sigma}
- g_{\mu \rho} J_{\nu \sigma} + g_{ \mu \sigma} J_{\nu \rho} - g_{\nu
\sigma} J_{\mu \rho} \right) \EQN corres-1;c \cr}
$$
showing that the algebra in momentum space is completely consistent with
that of
the usual Poincar\'e group.   The Casimir invariant operators of the
Poincar\'e group
are maintained as shown by virtue of equation~\Ep{corres-1;b}  and  it is
easy to see
that~$W_ \mu W^\mu$, where~$W_ \mu = -{1 \over 2} \epsilon_{\mu \nu \rho
\sigma} J^{\nu \rho} p^{\sigma}$
is the Pauli--Lubanski pseudovector,  must remain the second Casimir
invariant.
We are therefore free to characterise particles in terms of mass and
spin, just
as in standard field theory.  We also have
$$
\EQNalign{
[x_ \mu, J_{\rho \sigma}] &= i(g_{\mu \rho} x_ \sigma - g_{\mu \sigma} x_
\rho) \EQN corres-2;a  \cr
[x^2, J_{\rho \sigma}] &=0 \EQN corres-2;b \cr
}
$$
The Lorentz invariant nature of the operator-based space-time is made
apparent
in~\Ep{corres-2;b}
since the light-cone operator $x^2$ commutes with the generators of
Lorentz transformations.


\section{Massless Scalar Field in  Snyder Space-Time}

\subsection{The Geodesic Ansatz and Green's Function}

The usual operations of (coordinate space) analysis are not available in
quantised
spacetime since the spectra of the position operators is discrete and
also cannot be simulataneously diagonalized because of the
non-commuting nature of these operators.
Thus it is necessary to formulate field theory in momentum space where
the momenta
retain a continous spectrum.  As a first step it is necessary to
construct the propagator (two-point function)
in momentum space  using the momentum space representation of the
coordinate operators.

To determine the propagator in the quantised spacetime, first consider the
fundamental Green functions for massless (free) scalar fields in ordinary
quantum field theory.
The  different Green's functions
(retarded, advanced, Feynman, etc.) for a massless scalar field can all
be represented by the formula
$$
G(x, x') = {1 \over{(2 \pi)^n}}\int\!\!\d^n p \, \e^{-i p (x-x')} \tilde
G(p) \EQN{(4--6)}
$$
where~$\tilde G(p) = 1/p^2$ plus the  prescription made for the avoidance
of the
singularity in the complex momentum plane  (see Birrell and
Davies\cite{birdav}).
Carrying out the integration explicitly for the retarded and advanced
Green's functions, we obtain in 4-D
$$
G\!\!_{\textstyle{\matrix{ \hbox{\sevenrm ret}
\cr\noalign{\vskip-10pt}\hbox{\sevenrm adv}\cr}}}\!(x)
= {1 \over {2 \pi}} \theta(\pm x_0)\, \delta(x^2) \EQN{(4--10)}
$$
where~$\theta(x)$ is the unit step function.  Thus support is
concentrated on the
forward light-cone in the case of the retarded Green's function, and
concentrated on
the backward light-cone for the advanced Green's function.  We can thus
see that the
product of the light cone  geodesic with the retarded or advanced Green's
functions~(in 4-D) vanishes,
$$
x^2\,G\!\!_{\textstyle{\matrix{ \hbox{\sevenrm ret}
\cr\noalign{\vskip-10pt}\hbox{\sevenrm adv}\cr}}}\!(x) = 0. \EQN{(4--11)}
$$

To extend this idea to the quantised space time, consider the state
$\vert G\rangle$
representing the Green functions.  We then begin with a (complete) set of
states
$\vert x\rangle\equiv \vert \lambda_z, \lambda_1, \lambda_2\rangle$
which are eigenstates of the set of commuting operators $x_3$
(i.e. one of the four coordinate operators)
and the two casimir invariants
$C_1$ and $C_2$ from \Ep{cas}.  In particular, $\hat x_3\vert x\rangle
=\lambda_z\vert x\rangle$
where $\hat x_3$ denotes the operator and $\lambda_z$ denotes the
eigenvalue.
Then it is evident that $\langle x\vert G\rangle$ is a c-number function
of the eigenvalues
$\lambda_z,~\lambda_1,~\lambda_2$.
The ansatz is then made that the expectation value of the light cone
operator with the Green's
functions vanishes in the quantised space-time.  With this ansatz, we have
the fundamental constraint $\langle x \vert \hat x^2G\rangle =0$, leading
to
$$
\eqalign{
 \langle x \big\vert \hat x^2 \big
\vert G \rangle
& = \int\!\! {{\d^4 p} \over{ f(p)}} \langle x \big\vert p \rangle\,
\langle p \big\vert \hat x^2
\big\vert G \rangle\cr
 &= \int\!\!{{\d^4 p} \over{ f(p)}} \langle x \big\vert p \rangle\,
\tilde x^2 \langle p
\big\vert G \rangle
 = \int\!\!{{\d^4 p} \over{ f(p)}} \langle x \big\vert p \rangle\, \tilde
x^2
\tilde G(p)
 = 0 } \EQN ansatz1
$$
where~$\hat x^2$ represents the
Lorentz invariant
light-cone operator, ~$\tilde x^2$ is the momentum-space
representation of this operator, and the momentum space Green function
$\tilde G(p)=\langle p\vert G\rangle $ is a function of the continuous
momentum
eigenvalues $p_\mu$.

The function~$f(p)$ is a measure required for symmetries
of~\Ep{quadform}~and is also necessary for the
operator~$\tilde x$ to be Hermitian in the space of functions of
momentum\cite{snyder1}.
To determine the measure, consider the following analysis.
$$
\eqalign{
\langle \varphi \vert  x_ \mu \vert \Psi \rangle & = \int\!\!{{\d^4p}
\over{f(p)}
} \varphi^\ast(p) \tilde x_ \mu \Psi(p)\cr
& =\int\!\!{{\d^4p} \over{f(p)}}\, \varphi^\ast i (\partial_ \mu  - a^2
p_ \mu p
 \cdot \partial ) \Psi \cr
& = \int\!\!{{\d^4p} \over{f(p)}}\, (\tilde x_ \mu \Phi)^\ast \Psi
- i \int\!\!\d^4p\, \varphi^\ast \left( (\partial_ \mu {1 \over{f(p)}}) -
\partial_ \rho \left({{a^2 p_ \mu p^\rho} \over{f(p)}} \right) \right)
\Psi
\cr}\EQN measurederive
$$
The operator will be Hermitian if $f(p)$ can be found such that the
second term
in the last line above vanishes.  Thus we obtain a differential equation
for $f(p)$,
$$
i\left( \partial_ \mu - a^2 p_ \mu \left(p \cdot \partial - 5 \right)
\right) f(p) = 0 .\EQN measurediff
$$
with solution
$$
f(p) = (1 - a^2 p^2)^{5/2} \EQN measure
$$
This result is  notably identical to the volume element
of the curved momentum space considered by Gol'fand\cite{golfand1}.

Before solving the defining equation of free Green's functions
$\tilde x^2 \tilde G(p) = 0$ in the quantised space-time,
it helps to further develop the geodesic analogy between quantised and
standard space-time.
Consider the product~$x^2 G(x)$, where~$G(x)$ is the general Green's
function
defined in equation~\Ep{(4--6)}. In a manner exactly analogous to
\Ep{ansatz1}
we take~$x_\mu=i\frac{\partial}{\partial p^\mu}$ as an operator and with
simple manipulation we can show that
 $$
 x^2 G(x) = {-1 \over {(2 \pi)^n}}\int\!\!\d^n p\, \e^{-i p x} \square\,
\tilde G(p) \EQN{(4--12)}
 $$
and further,  using the definition of the general propagator,  we can
easily see that
 $$
 \square\, \tilde G(p) = \square\, {1 \over{p^2 }} = {{8 - 2n}
\over{(p^2)^2}} = 0 , \qquad p^2 \neq 0 \EQN{(4--14)}
 $$
 which vanishes when the number of dimensions~$n$ is equal to~$4$.
Therefore,~$1/p^2$
 is a solution of the differential equation~$\square\, {\tilde G} = 0$,
in four dimensions.
 Thus in standard  massless scalar quantum field theory we can write
$$
\int\!\!{{\d^4p}\over{(2\pi)^4}} \e^{ipx} x^2 \tilde G(p) = 0, \EQN{(4--15)}
$$
where~$x^2$ is treated as an operator.

Therefore, in  quantised space-time, the equation of free
Green's functions is obtained from \Ep{ansatz1} as
$$
 \int\!\!{{\d^4p}\over{(2 \pi)^4}} \, {{\langle x \big\vert p
\rangle}\over{(1 - a^2 p^2)^{5 / 2}}}
 \tilde x^2\tilde G(p) = 0. \EQN{(4--17)}
$$
Note for clarity  that~$\tilde x^2$ is  not acting  on the  measure,
since it is considered to
multiply the operator from the left.  Thus~$ \tilde x^2 {\tilde G}(p) =
0$ in momentum space.
Writing the coordinate operators explicitly then gives the differential
equation
$$
i( \partial_ \mu - a^2 p_ \mu p^\rho \partial_ \rho) i(\partial^\mu - a^2
p^\mu p_\sigma \partial^ \sigma) \tilde G(p) = 0 \EQN{(4--19)}
$$

 From symmetry considerations we expect that~$\tilde G$ will be a
function of $p^2$.
The partial differential equation is then  converted to the ordinary equation
$$
a^2 \left( 4 y (1 - y ) {{\d^2} \over{\d y^2}} - 2 (3 y - 4) { \d \over{
\d y}} \right) \tilde G(y) = 0 \EQN{(4--23)}
$$
where~$y= a^2p^2$.  This equation can be solved for all real values of
$y$ subject to the
conditions that $\tilde G(y)\rightarrow 0$ as $y\rightarrow
\pm\infty,~y\tilde G(y)\rightarrow a^2$ as
$y\rightarrow 0$.  This latter condition ensures that $\tilde
G(p^2)\rightarrow 1/p^2$ as
$a\rightarrow 0$, corresponding to the usual propagator for massless scalar
fields.  If $y<0$ the solution to \Ep{(4--23)}
subject to the above conditions is given by
$$
\tilde G(p) =  {{\sqrt{1 - a^2 p^2}}\over{p^2} }- a^2 \coth^{-1}\! \left(
\sqrt{1 - a^2 p^2} \right) \EQN{(4--28)}
$$
We obtain a momentum-space representation of this propagator by
continuing this
solution to complex momentum, utilizing standard prescriptions for avoiding
singularities in the momentum plane:
$$
\cases{\hbox{Feynman}& $p^2 \to p^2 + i \epsilon$;\cr
\eqalign{&\hbox to46pt{retarded}\cr \noalign{\vskip-9pt}& \hbox
to46pt{advanced}\cr}
& $p^2 \to (p_0 \pm i \epsilon)^2 - {\bf p}^2$\cr}\EQN{(4--30)}
$$
Subject to these restrictions, we obtain the Fourier representation
$$
G(x) = \int\!\!{{\d^4p}\over{(2 \pi)^4}}\, {{\langle x \big\vert p
\rangle} \over{(1 - a^2 p^2)^{5/2}}}
\left({{ \sqrt{1 - a^2 p^2}} \over{p^2}} - a^2 \coth^{-1}\! \left(\sqrt{1
- a^2 p^2} \right) \right) \EQN propagator
$$
which yields the usual scalar field Feynman propagator
$$
G_F(p)=\int\!\!{{\d^4p}\over{(2 \pi)^4}} \,{{\e^{-ipx}} \over{p^2}} \EQN
stdprop
$$
in the $a\rightarrow 0$ limit.
To understand this, note that the transformation Kernel $\langle x\vert
p\rangle$ must
satisfy  relations following from the property of the state $\vert
x\rangle$ as
 eigenstates of the operators $\hat x_3$, $C_1$ and $C_2$.
$$
\EQNalign{
\langle x\vert \hat x_3\vert p\rangle &= \lambda_z\langle x\vert
p\rangle =
-\left(\partial_{p_z}-a^2p_z p\cdot\partial\right)\langle x\vert p\rangle
\EQN kerneldef;a\cr
\langle x\vert C_1\vert p\rangle &=\lambda_1\langle x\vert p\rangle\EQN
kerneldef;b\cr
\langle x\vert C_2\vert p\rangle &=\lambda_2\langle x\vert p\rangle\EQN
kerneldef;c
}
$$
Then when $a\rightarrow 0$, the casimirs become $C_1\rightarrow \hat
x^2$, $C_2\rightarrow 0$
and $\tilde x_\mu\rightarrow \partial/\partial p_\mu$
which gives
$\langle x\vert p\rangle\rightarrow e^{-ip\cdot x}$
in the $a\rightarrow 0$ limit.

As usual the free Green function is associated with a time ordered product:
$$
\langle 0 \vert T\bigl(\varphi(x_1) \varphi(x_2)\bigr) \vert 0 \rangle =
i G(x_1 - x_2) \EQN{(4--31)}
$$
with~$G(x)$ given by~\Ep{propagator}, with the time ordered product
implying that
the Feynman prescription for the avoidance of singularities in the
complex momentum plane
is to be employed.  In this manner we have arrived at a definition of the
propagator
of our theory, through an ansatz which involves the geodesic equation of
the light cone.
An alternative propagator, based on the solution of \Ep{(4--23)} for all
real $y$, is
discussed in the appendix.

It is important that the propagator has analyticity properties consistent
with causality,
and a singularity and branch cut structure which does not prevent one
from making the usual
Wick rotation into Euclidean space.  Also, since the integral measure is
closely
associated with the propagator, it must be included in this analysis.

It is evident that the integrand of \Ep{propagator} with $p^2\rightarrow
p^2+i\epsilon$
has singularities in the complex $p_0$ plane located at
$$
p_0=\pm(|\vec p|-i\epsilon)\qquad\quad p_0=\pm(\sqrt{\vec
p^{\,2}+1/a^2}-i\epsilon )
\EQN wick1
$$
The second term of \Ep{propagator} also has branch cuts from
both the integration measure and the inverse hyperbolic function.
The integration-measure branch cut  occurs when $[1-a^2(p^2+i\epsilon)]$
is negative and real, corresponding to two branch cuts in the complex $p_0$
plane.
$$
\EQNalign
{
Im(p_0)&=-i\epsilon \,,\quad Re(p_0)>\sqrt{\vec p^{\,2}+1/a^2}\EQN wick2;a\cr
Im(p_0)&=+i\epsilon\,,\quad
Re(p_0)<-\sqrt{\vec p^{\,2}+1/a^2}\EQN wick2;b\cr
}
$$
The branch cut in the inverse hyperbolic cotangent begins when its
argument is
real and less than one, corresponding to positive real values of
$y=a^2(p^2+i\epsilon)$.  In the complex $p_0$ plane, this cut
adds the following two segments to the branch cuts of\Ep{wick2}:
$$
\EQNalign
{
Im(p_0)&=-i\epsilon \,,\quad \vert \vec p\vert <Re(p_0)<\sqrt{\vec
p^{\,2}+1/a^2}\EQN wick3;a\cr
Im(p_0)&=+i\epsilon\,,\quad -\sqrt{\vec p^{\,2}+1/a^2}<Re(p_0)<
-\vert \vec p\vert
\EQN wick3;b \cr
}
$$
As is evident from the above analysis, integration along the real $p_0$
axis can be performed along the Wick-rotated contour without
encountering
any singularities, consistent with the analyticity (and causality) properties
of the Feynman propagator expected from standard quantum field theory.

We reiterate that the first term in \Ep{propagator} contains no branch
cuts of its own,
in which case the branch structure of the entire propagator is dictated
by the second term.
In the complex $p^2$ plane, the integration-measure branch cut begins at
$p^2=1/a^2$ and extends to $+\infty$, indicative of particle creation
at
energies sufficiently large to probe the structure of the quantised
spacetime.
The branch segment \Ep{wick3}, which covers all positive values of $p^2$
below $1/a^2$, is proportional to $a^2$ and must be considered to be a
weak effect
of the quantised spacetime occurring at all energy scales in the theory.

\subsection{The Perturbation Expansion}

To construct the perturbative expansion of field
theory in quantised space-time, we first define the extended
fields~$\varphi_e$ as superpositions of extended plane wave
solutions\cite{nams2}
$$
{{G(\hat p)^{-1/2}} \over{\sqrt{\hat p^2}}} \braket p x \EQN fielddef
$$
where~$G$ is given by~\Ep{propagator}, and~$\hat p$ is formally the
momentum operator
in coordinate space. One can then write down a Lagrangian density
in terms of these extended fields in the usual way
$$
{\scr L} = {1 \over 2}\varphi_e G(\hat p)^{-1} \varphi_e +
{\scr L}_{\rm int}(\varphi_e).
$$
Since commutators of field operators remain c-numbers, and  time- and
normal-ordering
remain well-defined since time and momentum space are continuous, the
S-matrix
can be written in terms of these extended field objects as
$$
S =\, :\exp \left\{\sum_z M(z) {\varphi_e}_{\rm in}(z)
\, G(z)^{-1}{{\delta}
\over{\delta J(z)}}\right\}: Z[J]\vbigl{\vert}{0.6in}_{J = 0} \EQN s-matrix
$$
where~$Z$ is the generating functional of Green functions,~$J$ is the
usual Schwinger field source and~$M(z)$ is a possible coordinate space
measure, analogous to the momentum space measure discussed previously.
The sum/integral over the eigenvalues $\lambda_z,~\lambda_1,\lambda_2$ of
the states
$\vert x\rangle$ are heuristically denoted by $\sum_z\,$.
Thus
standard perturbation theory is valid in terms of the
extended fields~$\varphi_e$.

Calculations in field theory are generally carried out in momentum space due
to the difficulty of working in coordinate space.  Here we are lacking any
definite method of  computation in
coordinate space and rely completely on the momentum space formulation of
the theory.
The formal transformation of the S-matrix elements into momentum
space is straightforward, provided that we have
$$
\EQNalign{
\sum_x M(x) \braket p x \braket x q &= \delta (p - q) \EQN gotta;a \cr
\braket {x - y} p & = \braket x p \braket p y .\EQN gotta;b\cr}
$$
Equation \Ep{gotta;b} is easy to show from the defining equation
\Ep{kerneldef;a}
provided we make the natural identification of the eigenvalue
$\lambda_x-\lambda_y$ with the eigenvalue $\lambda_{x-y}$.
However, \Ep{gotta;a} is more problematic.  Lacking definite solutions of
\Ep{kerneldef}, except in the~$a \to 0$ limit, we will assume that this
condition
is satisfied for~$a \neq 0$.  This amounts to the assumption of
conservation of momentum.


\section{One-Loop Calculations}

It is now possible to formulate the one-loop processes in~$g \varphi^4$
theory.  Using~\Ep{s-matrix} we have the two point function as
$$
\eqalign{
&\bra O T\bigl(\varphi_e (x) \varphi_e (0) \bigr) \ket O =
D(x)=\braket x D\cr
&\quad= g \sum_y M(y) \bra O T \bigl(\varphi_e (x) \varphi_e(y)\bigr)
\ket O \bra O \varphi_e (y) \varphi_e (y)  \ket O \bra O T
\bigl(\varphi_e (y) \varphi_e (0) \bigr) \ket O\cr
&\quad= g \sum_y M(y) G(x-y) G(0) G(y) }\EQN one-loop
$$
then
$$
\eqalign{
D(p)&=\braket p D =\sum_x M(x) \braket p x \braket x D \cr
&= g G(0) \sum_x M(x) \braket x p
\sum_y M(y) G(y-x) G(y)\cr
&=g G(0) \sum_x M(x) \braket x p  \sum_y M(y)
\int\!\!{{\d^4 l} \over{M(l)}} \int\!\!{{\d^4 s} \over{M(s)}} \braket l
{y-x} \braket s{y} \tilde G(l) \tilde G(s)\cr
&= g G(0) \sum_x \sum_y \int\!\!{{\d^4 l} \over{f(l)}}
\int\!\!{{\d^4 s} \over{f(s)}} M(x) \braket l x \braket x p M(y) \braket
l y \braket y s \tilde G(l) \tilde G(s) \cr
& =g {{\tilde G (p)} \over{f(p)}}\, G (0)\,{{\tilde G (p)} \over{f(p)}}
\,\,
= \,
\frac{\tilde G(p)}{f(p)}\quad
g \int\!\!{{\d^4k} \over{(2 \pi)^4}} {{\tilde G(k)} \over{f(k)}}
\quad \frac{\tilde G(p)}{f(p)}
\cr }
\EQN stdtwop
$$
where we have used~\Ep{gotta}. The external propagators and their associated
measures are, of course, removed to give the truncated Green function.
The four point function can be handled
the same way.

\subsection{ Particle Self-Energy to One Loop in $g \varphi^4$}

\figure{selfenergy}
\infiglist{First order self energy diagram.}
\readfigfile{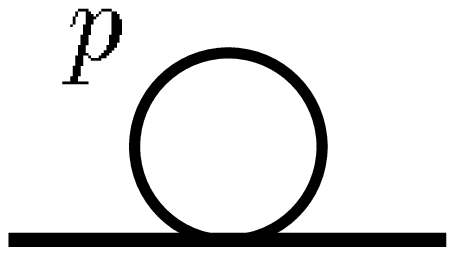}{1.5}
\Caption
First order self energy diagram.
\endCaption
\endfigure

The application of the Feynman rules to the vacuum diagram
\Fig{selfenergy} gives us the following integral:
$$
g\int\!\!{{\d^4p}\over{(2 \pi)^4}} \,  {1 \over{(1 - i \epsilon - a^2
p^2)^{5/2}}}
\left( {{-\sqrt{1 - i \epsilon - a^2 p^2}} \over{p^2 + i \epsilon}} + a^2
\coth^{-1} \!
\left( \sqrt{1 - i \epsilon - a^2 p^2}\, \right) \right).\EQN{(4--34)}
$$
The result for the first order contribution to the self-energy
is\Footnote{\dag}{MAPLE was found useful for several computations.
Symbolic Computation Group,
MAPLE (a computer program) version V, Computer Science Dept., University
of Waterloo, Waterloo, Ontario, 1990}
$$
-i \Sigma(0) = G(0) = {{ig } \over{24 \pi^2 a^2}} \left[ 1 + 2 \log 2
\right] \EQN{(4--41)}
$$
Note that~$\lim_{a \to 0} G(0)$ is unbounded, as these integrals normally
are in field theory
without regularisation or renormalisation.  The quantised space-time
thus regulates the UV divergences of the two-point function.

To determine how the self energy shifts the poles in the propagator away
from the
bare values, we write the complete propagator~$G_c$  as
$$
iG_c(p) = iG(p)\left({1 \over{1 + i \Sigma(p) iG(p)}}\right) \EQN{(4--42)}
$$
Where~$G(p)$ is the momentum space propagator of the present theory.
Putting
in the complete expression for~$G(p)$ gives
$$
iG_c(p) = {{i\left(- \sqrt{1-a^2p^2} +
a^2p^2\coth^{-1}\!\left(\sqrt{1-a^2p^2}\, \right)
\right)}\over{p^2(1-a^2p^2)^{5/2}
- {{\Sigma(p)}}\left(-\sqrt{1-a^2p^2} + a^2p^2
\coth^{-1}\!\left(\sqrt{1-a^2p^2}\, \right) \right)}} \EQN{(4--43)}
$$

Nothing essential has been altered.  The  points~$p^2 = 0$
and~$p^2=1/a^2$ remain branch points,
and thus the same resonances exist here as in the bare propagator.  The
physics is not altered in any essential way.

This is analogous to standard   $g \varphi^4$
field theory in dimensional regularisation, where no shifting of the mass
shell
for a massless particle occurs to one-loop because the massless tadpoles
are zero.
In this case, however, rather than dealing with a
mass shell, we have a resonance resulting from the quantised
space-time.

The vacuum diagram considered here is known to be the highest-degree
primitive
divergence in~$g \varphi^4$ field theory.   Since we have shown that this
theory renders this diagram finite, we can thus argue that all diagrams
in~$g \varphi^4$
are rendered ultraviolet finite by this approach, thus making the entire
theory ultraviolet finite.   The asymptotic behaviour of the propagator
in equation~\Ep{propagator} is~$\sim 1/p^2$. This takes into account  the
four powers
of the integration variable in the numerator as well as the integral
measure.  We
thus have, in the worst case, quadratic ultraviolet {\it convergence.}

The infrared behaviour is also convergent, for this self-energy diagram,
since a sufficient number of
powers of the loop momentum in the numerator  force the inverse
hyperbolic cotangent and the~$1/p^2$ behaviour of the first term to zero
as~$p^2 \to 0$.

\subsection{The Four-Point Function}

\figure{fourpoint}
\infiglist{Four-point diagram.}
\readfigfile{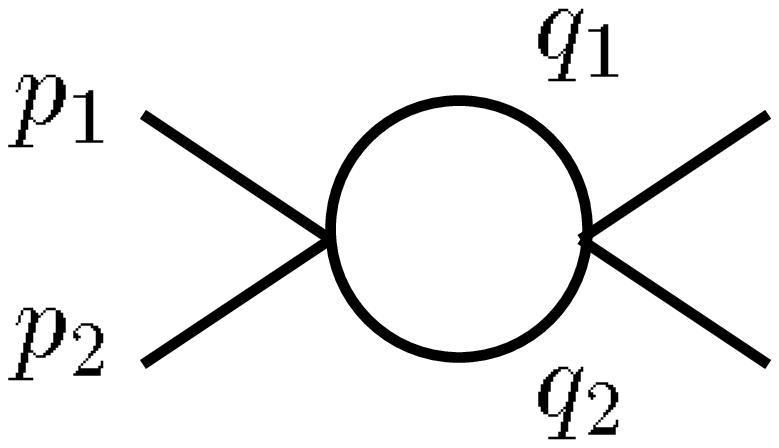}{1.5}
\Caption
Four-point diagram.
\endCaption
\endfigure

Application of the Feynman rules to the diagram of~\Fig{fourpoint}, which
represents
one of three possible channels for this process, results in the integral
$$
\EQNalign{
{{g^2} \over {(2 \pi)^4}} \int\!\!&{{\d^4q} \over{(1 -a^2q^2)^{5/2}\, (1
- a^2(p-q)^2)^{5/2}}}\,
\left( {{\sqrt{1 - a^2 q^2}} \over{q^2}} - a^2 \coth^{-1} \left( \sqrt{1
-a^2q^2} \right) \right)\cr &
\qquad\times \, \left({{\sqrt{1 -a^2(p-q)^2}} \over{(p-q)^2}} - a^2
\coth^{-1}  \left( \sqrt{1 -a^2(p-q)^2} \right) \right), \EQN fourpointint}
$$
each of the channels is calculated through a similar integration.

Expansion of \Ep{fourpointint} using $\coth^{-1} (x) = x \int_0^1{{\d u}
\over{x^2 - u^2}}$ gives us the four integrals
$$
\eqalign{
I_1 = &{{g^2 \mu^8} \over{(2 \pi)^4}} \int\!\!{{\d^4q} \over{q^2\,(\mu^2
- q^2)^2\, (p-q)^2 \,(\mu^2 - (p-q)^2)^2}}\cr
I_2 = & {{-g^2 \mu^8} \over{(2 \pi)^4}} \int_0^1\!\!\d u \int\!\!{{\d^4
q} \over{q^2\, (\mu^2-q^2)^2\, (\mu^2- (p-q)^2)^{2}\, ( \sigma^2 -
(p-q)^2)}}\cr
I_3 = &{{-g^2 \mu^8} \over{(2 \pi)^4}} \int_0^1\!\!\d u\int\!\!{{\d^4 q}
\over{(p-q)^2\, (\mu^2-(p-q)^2)^2\, (\mu^2-q^2)^{2}\,(\sigma^2 - q^2)}} \cr
I_4 = & {{g^2 \mu^8} \over{(2 \pi)^4}} \int_0^1\!\!\d u \int_0^1\!\! \d v
\int\!\!{{\d^4 q}
\over{(\mu^2 - q^2)^2\, (\mu^2 - (p-q)^2 )^2\, (\sigma^2 - q^2)\, (\xi^2
- (p-q)^2)}}\cr}
 \EQN fourpointint-2
$$
where we have put~$\mu^2 = a^{-2}$, $\sigma^2 = \mu^2 (1- u^2)$
and~$\xi^2 = \mu^2 (1- v^2)$.
It is obvious that~$I_3 = I_2$ with change of variable. Integrals~$I_1$,
$I_2$,
and~$I_4$ can be computed through decomposition into partial fractions,
for example
$$
{1 \over{q^2 (\mu^2 - q^2)}} = {1 \over{\mu^4 q^2}} +
{1 \over {\mu^4 (\mu^2 - q^2)}} + {1 \over{\mu^2 (\mu^2 - q^2)}},\EQN parfrac
$$
thus for~$I_4$ we obtain nine integrals of the form
$$
{{g^2 } \over{(2 \pi)^4}} \int_0^1\!\!{{\d u} \over{u^4}}
\int_0^1\!\!{{\d v}
\over{v^4}}\int\!\!{{\d^4q} \over{\bigl(\mu^2 - q^2\bigr) \bigl(\mu^2 -
(p-q)^2\bigr)}}\EQN exint
$$
each of which is rewritten in a  dimensionally regularised form, ie., for
\Ep{exint}
$$
{{g^2} \over{\nu^{2 \epsilon}}}\int_0^1\!\!{{\d u} \over{u^4}}
\int_0^1\!\!{{\d v} \over{v^4}} \int\!\!{{\d^Dq} \over{(2 \pi)^D}}
{1 \over{\bigl(q^2 - \mu^2 + i \eta\bigr) \bigl( (p-q)^2 - \mu^2 + i \eta
\bigr)}} \EQN dimint
$$
and the integral obtained by referring to existing compilations of such
integrals
\cite{PandT}\cite{kreimer}.   This procedure is carried out for
each of~$I_1$, $I_2$, and~$I_4$.  Dimensional regularisation is simply an
intermediate step in this calculation.

It is found that for each integral  the divergent parts of the sub-integrals
vanish for~$D \to 4$, as one would expect for a finite integral.  A
certain question
remains regarding the~IR finiteness of the integrals over the Feynman
parameters~$u$
and~$v$ in the case of~$I_2$ (and~$I_3$) and~$I_4$. One can see that in
the example, \Ep{dimint},
these integrals are divergent. The final result for the
four-point function is sufficiently complex that it precludes a direct
analysis.
We can, however, expand each of the nine integrals of~$I_2$ and~$I_4$ in a
power series in~$u$ and~$v$ to determine the behaviour as~$u$ and~$v$
approach
zero.  For each integral the divergent terms in the series completely
vanish.
We can therefore conclude that the behaviour of the four point function
near~$u=v=0$ is finite, and therefore there are no~IR divergences.

\subsection{ Comparison of Quantised-Space Propagators}

In this section we will compare the propagator of the present theory,
obtained
in section~3.2, to that obtained in the related theory of
Namsrai\cite{nams1}.
We wish to note the dissimilarity in the convergence behaviour of each of
these
propagators as a function of increasing~(Euclidean) momentum.

 From Namsrai~(op cit) we have the massless free scalar propagator  of
the stochastic theory as
$$
N(p) = {1 \over{p^2 \cosh^2 \left(a \sqrt{-p^2} \right)}} \EQN{(4--70)}
$$

In the energy regime~$p^2 < 1/a^2$, the two propagators are very nearly
identical, however their asymptotic behaviour is quite different.  In the
case of~\Ep{(4--70)}
we have exponential suppression, whereas for~\Ep{propagator} there is
power-law decay. This indicates that the stochastic and
non-stochastic formulations are distinct.

\section{Conclusions}

A quantum scalar field is constructed in a space-time generated by an
operator
algebra originally proposed by Snyder\cite{snyder1}.   This algebra
introduces
a fundamental length scale, leading to  quantised spatial and continuous
temporal
coordinates .  The Snyder algebra contains the  Poincar\'e group with its
Casimir
invariants, allowing particles to be described in terms of spin and mass.

The retarded and advanced Green's functions in standard scalar quantum
field theory
have support on the light cone in the massless case.  This is used to
construct
Green's functions for the massless scalar field in the quantised
space-time directly in
momentum space, through the momentum space ansatz  that an exactly
analogous equation
holds in the quantised space-time.  The measure that must be introduced
into the theory
from the symmetries of the space-time algebra not only ensures the
space-time coordinate operators are Hermitian,
but also corresponds to the volume element in the constant-curvature
momentum space considered by Gol'fand.

We find  that the propagator has a branch structure indicating the presence
of resonance thresholds. There are two main effects.  The first results
in the
creation of particles of mass~$ m = 1/a$ at energies sufficient to probe
the structure
of the quantised space-time.  The other is a much weaker threshold
at~$p^2 = 0$, proportional to~$a^2$, which is an effect due to the
quantised space-time throughout the energy range of the theory.

The self-energy and the four-point function  in~$g \varphi^4$ were
calculated
to one-loop and
shown to be ultraviolet finite,
demonstrating that the fundamental length scale of the space-time algebra
regulates ultraviolet divergences.   These one-loop processes were also
found to be infrared finite.

Throughout the calculations undertaken, it is emphasised that in the
limit as the
fundamental length of the space-time vanishes, all results make sense in
terms of
standard quantum field theory. In particular, the self-energy and four
point function are singular as~$a \to 0$
and the free propagator~\Ep{propagator} reduces to its continuous
space-time counterpart.

Also worthy of note is the position-momentum uncertainty relation derivable
from equation~\Ep{commute-set}.  So-called generalised uncertainty
relations have been of recent interest in superstring theory and quantum
gravity\cite{net}\cite{genun1} and seem to arise naturally
from these theories.  It is possible that quantised space-times and more
general uncertainty relations imply one another.

It is hoped that this non-stochastic approach to massless QFT in a
quantised space-time can be extended to massive fields in future research.

\nosechead{Acknowledgementts}

We are grateful for the financial support of the Natural Sciences and
Engineering Research Council of Canada.
\bigskip
\vfill\eject
\nosechead{Appendix}
\nosechead{Alternate Solution for the Propagator}

There exists a second solution to the propagator equation,
\Ep{propagator}.
In the course of solving~\Ep{(4--23)} we obtain the solution in the form
of the integral
$$
{\tilde G}(y) = C \int\!\!\d y\, {{{\sqrt{\mathstrut\lower2pt\hbox{$\vert
1-y \vert$}}}\over{y^2}}}, \EQN{(4--26)}
$$
where~$y= a^2 p^2$.
This equation can be solved for all real values of $y$, subject to the
conditions
stated prior to \Ep{(4--28)}, as well as the further assumption that purely
$1/p^2$ behaviour occurs in the immediate neighbourhood of $p^2=0$.
( $1/p^2+ constant$ is in principle compatible with the condition
$y\tilde G(y)\rightarrow a^2$ as $y\rightarrow 0^+$).  The integrand of
the
resulting solution
$$
G(x) = \int\!\!{{\d^4p\,{\,{\braket x p}}}\over{(2 \pi)^4 (1 - a^2
p^2)^{5/2}}}
\cases{ {{ \sqrt{1 - a^2 p^2}} \over{p^2}} - a^2 \coth^{-1}\!
\left(\sqrt{1 - a^2 p^2} \right)
& $p^2 < 0$ \cr {{ \sqrt{1 - a^2 p^2}} \over{p^2}} - a^2 \tanh^{-1}\!
\left(\sqrt{1 - a^2 p^2} \right)  & $0 \le a^2p^2 < 1$ \cr
\hskip90pt 0 & $a^2p^2 \ge 1$\cr} \EQN alt-prop
$$
is defined {\sl only} for real values of $p^2$.  Thus the $p_0$
integration
in $d^4p$ must be performed over the real $p_0$ axis; a Wick rotation is
not permitted
because the solution does not have a unique analytic continuation
for complex $p_0$.  An interesting feature of \Ep{alt-prop} is the occurence
of a cut-off in the invariant mass.  Thus particles more massive that
$1/a^2$
cannot occur, consistent with the explicit assumption of others working
in
curved momentum spaces \cite{golfand1}.

\vfill\eject
\nosechead{References}
\ListReferences
\PrintFigures
\vfill\supereject

\bye